\documentclass[showpacs,showkeys,preprint]{revtex4}
\usepackage{graphicx}
\usepackage{dcolumn}
\usepackage{bm}
\usepackage{amsmath}
\usepackage{amssymb}
\usepackage{natbib}
\usepackage{subfig}

\begin{document}

\title{Rain  model for the transmission spectra of one dimensional disorder system }

\author{Zheng Liu }
\author{Xunya Jiang }
\email{xyjiang@mail.sim.ac.cn}
\affiliation{State Key Lab of
Shanghai Institute of Microsystem and Information Technology, CAS,\\
Functional Materials for Informatics, Shanghai 200050, China}

\date{\today}

\begin{abstract}
We imitate the spectrum character of one-dimensional disorder system with our new rain model. It
has been shown that the transmission spectrum can be approximately characterized  by the model,
which include some coupled lorentzian transmission peaks on the spectrum and embody the coupling of
the local modes in the disorder system.

\end{abstract}

\maketitle

The transport of optical waves through disordered systems can exhibit remarkable interference
effects, in analogy with the transport of electrons in solids. This leads to interesting optical
phenomena\cite{Ping} of which the most surprising is that of Anderson localization of
light\cite{John}. The observation of this extraordinary effect in three dimensional disordered
optical systems requires very strong scattering that can be achieved only in selected
materials\cite{three}. For lower dimensional systems, however, the situation is different. In one
and two dimensional disordered systems, localization can always be reached for a sufficiently large
sample size \cite{onetwo} However, not all modes are exponentially localized in 1D random systems,
even though Anderson localization occurs. Pendry\cite{jpc,jbad} and Tartakovskii
\cite{Tartakovskii} predicted that even in one dimensional(1D) localized systems nonlocalized modes
exist that extend over the sample via multiple resonances. These nonlocalized modes, called
necklace states(NS), have a transmission coefficient close to 1 and become extremely rare upon
increasing the sample thickness. Nevertheless, they dominate the average transmission coefficient,
even at large thickness. Evidence for the existence of necklace states was recently found in
time-resolved transmission experiments\cite{evid1}, and subsequently also observed in experiments
with microwaves\cite{evid2}. It is noted that these works about NS are mainly concentrated on the
one dimensional systems for its rather easily calculation and manufacture. The basic method and
ways is commonly to analyze the transmission spectra of the system whether
experimentally\cite{prlge1,prlge2} or theoretically\cite{prlns,theE}, since the transmission
spectra is essential for 1D disorder system  and reveal important formation of the system. For
example,when the phase of the transmission coefficient is the integral multiple of $2\pi$, it
corresponds to the order of NS\cite{theE} and the platforms in the spectrum usually can be used to
identify the occurrence of NS.   Since the transmission spectra directly implicate the intrinsic
properties of the system, It is reasonable that one can artificially construct transmission spectra
according to some aforehand rules inferred form the physical evaluation. Some statical properties
of physical variables of the system can be explored based on the constructed spectra. It is
equivalent  that one is inquiring interactively  the relations between the statical properties and
the possible physical mechanism. Based on the above idea, In this letter we present a model called
after rain model to imitate  the transmission spectrum of the one dimensional disorder system.
 Generally, the transmission spectra of localized 1D systems exhibit many randomly distributed
high-transmission peaks. These high-transmission peaks originate from resonances of the system with
localized modes\cite{Azbel} and result in big fluctuations in the transmission coefficient
T\cite{Ping,ja}. The localized modes decay exponentially and the ensemble average of over many
realizations of the disorder decays linearly with the sample thickness L\cite{jbad}.

 Our model is  as follows. Consider a one-dimensional disorder system with length L.  The localized
 states inhabiting in the system corresponding to the eigenfrequency $\omega_n$ usually have the
 form $E_n\propto e^{-|x-x_n|/\xi}$ except a oscillating factor, where $\xi$ is the localization
 length and $x_n$. For simplicity, omitting the fine structure of the  field for the moment we
 assume the field  localized at $x_n$  has simply the form of the envelope. Since the necklace
 states are usually  constructed by a series of  localized states  close to each other  spatially so that
 they seem like the necklace  and the resultant transmission  spectrum  manifest a high lorentzian peaks
 bundle. Therefor, first of all we should investigate what the transmission spectrum would be like if the
 fields in system are superposed  by multiple  localized states.  Here we have to use experiential
 conclusion that the transmission coefficient $t$ is expressed as $t=\frac{Min\{E_n(0),E_n(L)\}}{Max\{E_n(0),E_n(L)\}}$

 \begin{figure} \includegraphics[scale=0.45]{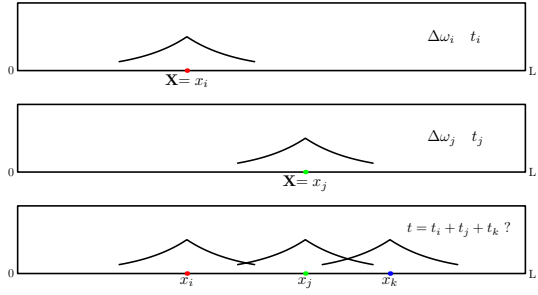}\caption{schematic figure of our model }\end{figure}

 With the above approximation and assuming  $x_n<L/2$, the transmission coefficient for a local state $x_n$
 is $t=\frac{E_n(L)}{E_n(0)}=e^{-L/\xi}e^{2x_n/\xi}$. For the field of $N$ coupled states
 $E_N(x)=\sum_{n=1}^{N}e^{-|x-x_n|/\xi}$, the corresponding transmission coefficient
  $t_N=\frac{E_N(L)}{E_N(0)}=\sum\limits_n^N t_n^{1/2}/\sum\limits_n^N t_n^{-1/2}$. Since the transmittance peaks
  in the bundle are lorentzian and has the form $T(\omega)\propto\frac{(\Delta\omega/2)^2}{(\omega-\omega_n)^2+(\Delta\omega/2)^2}$,
  where the center   frequency $\omega_n$  and $x_n$  are all the random variables and the halfwidth  $\Delta\omega$ can
  be determined by quality factor $Q$ with relation $Q=\frac{\omega_0}{\Delta\omega}$.
 According to the definition of $Q=\omega_0\frac{\mbox{stored energy}}{\mbox{power loss}}=\omega_0
 \frac{W_{store}}{P_{loss}} =\frac{\omega_0}{\Delta\omega}$\cite{jackson}, the halfwidth can be written as
  $\Delta\omega=\frac{P_{loss}}{W_{store}}$. In  one dimensional system how is  the quality factor  correlated with the transmittance?
  Since  quality factor $Q$  is the  measure  of the sharpness of response of the cavity to external excitation, it's reasonable that
  the system is  regarded  a cavity enclosed by the vacuum and the transmitted energy corresponds  the power loss.
 Here consider a imaginary pipe with unit area though the system. So the power loss can be
 expressed transmitted average energy current $\langle S\rangle_{x=L}=\frac{E(L)}{2Z_0}$ while the stored energy in the pipe $W_{store}=\int\langle U\rangle
\mbox{d}v=\int_0^L Re(\epsilon |E|^2)\mbox{d}x$.  It is found  from above results that
$\Delta\omega=\frac{E(L)^2}{2Z_0\epsilon\int E^2\mbox{d}x}=\frac{c E(L)^2}{2\epsilon_r\int
E^2\mbox{d}x}$,where $c$ is the light speed, $Z_0=\sqrt{\mu_0/\epsilon_0}$ is the vacuum impedance,
$\epsilon_r$ is the average relative permittivity of the system.   For the above field
$E_n(x)=e^{-|x-x_n|/\xi}$, we
obtain\begin{equation}\Delta\omega=-\frac{e^{4x_n/\xi}}{(e^{2L/\xi}+e^{4x_n/\xi}-2e^{2(L+x_n)/\xi}
)2\pi\epsilon_r }\frac{2\pi c}{\xi}\end{equation}

To imitate the spectrum of the necklace states we assume many lorentzian peaks  randomly fall on
the frequency axis with  random center frequency $\omega_n$, which is related with another spatial
random variable $x_m$ that is the center of local states.  For each $(\omega_n,x_m)$,  we first
construct the transmittance coefficient
\begin{equation}t_{nm}(\omega,\omega_n,x_m)=\frac{e^{(2x_m-L)/2\xi}(\Delta\omega/2)}{\sqrt{(\omega-\omega_n)^2+(\Delta\omega/2)^2}}\end{equation}

and the total transmittance spectrum can be obtained
\begin{equation}t_{total}(\omega)=\frac{\sum_{n,m}t_{nm}^{1/2}}{\sum_{n,m}t_{nm}^{-1/2}} \end{equation}

To explore the possible mechanism we introduce the repulsive interaction between these `frequency
raindrops' which are characterized by the corresponding  overlap integral between the two localized
states. Specifically speaking, for the states localized at $x_i,x_j$, the overlap integral is
$I_{i,j}=\int_{-\infty}^{\infty}e^{(-|x-x_i|-|x-x_j|)/\xi}dx=(\xi+\Delta x)e^{-\Delta x/\xi}$ and
the moduls square $\int_{-\infty}^{\infty}|E_(x)|^2 dx=\xi$.  In the rain process the frequency
points that satisfy the condition $I_{i,j}<\alpha\xi$  are permitted to be coupled into the
$t_{total}$, where parameter $\alpha$ control the coupling strength.  We call the process as two
order process since it only involves two coupled frequencies  and two localized states.  To check
the validity of the operation, we compared the correlation curves to frequency interval with those
obtained from the real one dimensional disorder system by transfer matrix method. The correlation
is defined as:\begin{widetext} \begin{equation}Cov(\Delta\omega)= \frac{\langle (\ln
T(\omega+\Delta\omega)-\langle \ln T(\omega+\Delta\omega)\rangle)(\ln T(\omega)-\langle \ln
T(\omega)\rangle)\rangle}{\sqrt{\langle (\ln T(\omega+\Delta\omega)^2-\langle \ln
T(\omega+\Delta\omega)\rangle^2)(\ln T(\omega)^2-\langle \ln
T(\omega)\rangle^2)\rangle}}\end{equation} \end{widetext}, where $\langle\cdot\rangle$ means the
average over all realizations of disorder.   For the two order process  we introduce a coupled mode
theory to phenomenologically characterize the effect on  the transmission coefficient by the fact
that mode coupling transform the two random frequencies $\omega_m,\omega_n$ into the renormalized
frequencies $\omega'_m,\omega'_n$:
\begin{eqnarray} \frac{d^2a_1}{dt^2}+\gamma_1\frac{da_1}{dt}+\omega_1 a_1&=&q_{12} a_2+e^{-i\omega'
t}\nonumber\\ \frac{d^2a_2}{dt^2}+\gamma_2\frac{da_2}{dt}+\omega_2 a_2&=& q_{21} a_1\end{eqnarray}

Where $a_1(t)=a_{10}e^{-i\omega_1t}$, $a_2(t)=a_{20}e^{-i\omega_2t}$ and  $a_{10}$, $a_{20}$,
$\omega'$ is to be determined. According to  the condition of  nonzero solution of $a_{10}$,
$a_{20}$,
\begin{equation}\label{newfr}(\omega_1^2-(i\gamma_1+\omega')\omega')(\omega_2^2-(i\gamma_2+\omega')\omega')- q_{12}
q_{21}=0\end{equation}

we can get two renormalized  peaks frequency $\omega'_1$, $\omega'_2$ with positive real parts,
where the factor $e^{-i\omega' t}$ describes the excitation of incident wave and  for simplicity
 the coupling coefficients $q_{12}=q_{21}=q\propto I_{i,j}$,  the damping factor  $\gamma_n\propto \Delta\omega$

  \setlength{\textfloatsep}{2pt plus 2pt minus 2pt}
 \begin{figure} \includegraphics[scale=0.4]{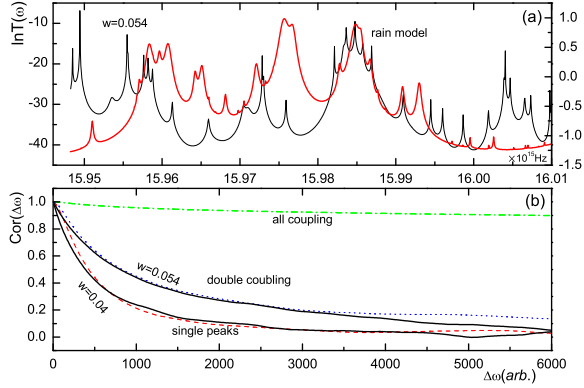}\caption{\scriptsize (a)$\mbox{ln}T(\omega)$ spectrum for
 Transfer method(solid) and rain model (dashed)  (b) Correlation for different frequency
interval  }\end{figure}
 In Fig.2 we compare the $\mbox{ln}T(\omega)$  and the correlation obtained by transfer matrix method for
 a $3000$ layers  disorder system with $d_1=200$nm, $\epsilon_A=1$, $d_2=100$nm, $\epsilon_B=2$ with the
 corresponding results obtained by the rain model.  In Fig.2(b) for one order approximation where
 lorenztian peaks are summed directly  the correlation by the rain model  is only in good agreement
 with the system with low random strength where $w=0.04$. However for higher random strength the one order
 approximation apparently fail in approximating the real disorder system.  For the real system our
 calculation show that when random strength $w$ grow  gradually  the correlation for a fixed frequency interval
 rise rapidly  and  descends  slowly after passing by a maximum, which is shown in Fig.3.
 Here  for our the calculated $3000$ layers' system  the critical random strength $w_c\simeq0.12$.
 It is found that when more and more  lorenztian peaks  are coupled into the transmission coefficient
 spectrum, which means the higher order processes, the whole correlation curve  also ascend rapidly as shown by the
 green dashed dotted line in Fig.2(b).   For a given random strength $w$  the correlation  of the  real system can
 be always approximated  by the rain model with specific number of  order process  because it can be  raised  consecutively
 with  the increment of number of  order.  For example, for $w=0.054$ rain model can give  a good  approximation
 with $\sim20000$ peaks  and $\alpha=0.5$, $L/\xi=8$ which is marked with blue dashed line in Fig.2(b).  The corresponding
 spectrum is shown in Fig.3(a).  It is noted  form  Fig.3(a) that  in the range $15.97\sim15.99$  the spectrums keep the
 close similarity and some double peaks structure begin to occur, which imply the thrown coupling double peaks play a
 important role in the similarity of correlation curve.

  \setlength{\textfloatsep}{2pt plus 2pt minus 2pt}
 \begin{figure} \includegraphics[scale=0.4]{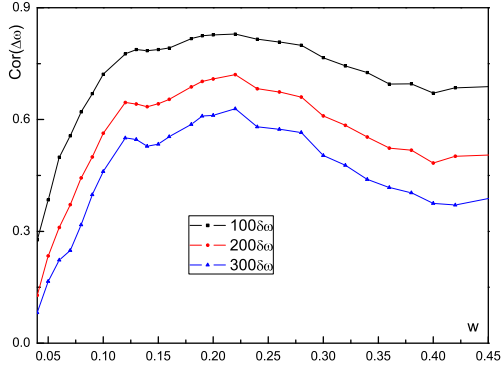}\caption{\scriptsize Correlation curve corresponding to different
  random strength $w$  for some fixed frequency intervals}\end{figure}

In conclusion, we present a model to simulate   the transmittance spectrum  of the one dimensional
disorder system.  The model describe the  construction of the spectrum based on a thought that many
lorenztian peaks can be  superposed in a way embodying  some interaction between these single
peaks. Furthermore, we introduce overlap integral between localized states and the coupling of
modes  to characterize  the interaction phenomenologically.  In the above  way lorenztian peaks
randomly drops on the frequency axis like raindrops  and for each peak  the characteristic
parameter is determined  by a localized states with a random position.  Here it involves two
aspects of  randomness:  frequency  and spatial domain. During the sampling  of the random
variables uniform distribution is used.  In the weak  disorder  the model can give rather better
approximation. Because the model associate  the form of transmittance spectrum with  the
interaction between single peaks and the position of the localized states  it provide a
intuitionistic tool for the subject that depends on the analysis  of the transmittance spectrum
such as the study of the necklace states.

\newpage

\end{document}